\documentclass[showkeys]{revtex4}
\usepackage{graphicx}
\usepackage{dcolumn}
\usepackage{amsmath}
\usepackage{color}
\usepackage{epstopdf}
\usepackage{float}
\usepackage{graphicx}
\usepackage{subfigure}
\usepackage[utf8x]{inputenc}
\definecolor{coolblack}{rgb}{0.0, 0.18, 0.39}
\definecolor{darkred}{rgb}{0.5,0,0}
\definecolor{darkgreen}{rgb}{0,0.5,0}
\definecolor{darkblue}{rgb}{0,0,0.5}
\definecolor{lapislazuli}{rgb}{0.15, 0.38, 0.61}
\definecolor{venetianred}{rgb}{0.78, 0.03, 0.08}
\definecolor{bleudefrance}{rgb}{0.19, 0.55, 0.91}
\definecolor{dogwoodrose}{rgb}{0.84, 0.09, 0.41}
\usepackage[linktocpage,colorlinks]{hyperref}
\hypersetup{colorlinks=true, citecolor=darkgreen, linkcolor=blue,urlcolor = darkblue}
\makeatletter
\def\btt#1{\texttt{\@backslashchar#1}}
\DeclareRobustCommand\bblash{\btt{\@backslashchar}} \makeatother

\RequirePackage{fix-cm}

\begin{document}
\title{Geodesic stability and quasinormal modes of non-commutative Schwarzschild black hole  employing Lyapunov exponent 
}
\author{Shobhit Giri $^{a}$}\email{shobhit6794@gmail.com}
\author{ Hemwati Nandan $^{a,b}$}\email{hnandan@associates.iucaa.in}
\author{Lokesh Kumar Joshi $^{c}$}\email{lokesh.joshe@gmail.com}
\author{Sunil D. Maharaj $^{d}$}\email{maharaj@ukzn.ac.za}
\affiliation{$^{a}$Department of Physics, Gurukula Kangri (Deemed to be University), Haridwar 249 404, Uttarakhand, India}
\affiliation{$^{b}$Center for Space Research, North-West University, Mahikeng 2745, South Africa}
\affiliation{$^{c}$ Department of Applied Science, Faculty of Engineering and Technology, Gurukula Kangri (Deemed to be University), Haridwar 249 404, Uttarakhand, India}
\affiliation{$^{d}$ Astrophysics Research Centre, School of Mathematics, Statastics and Computer Science, University of KwaZulu-Natal, Private Bag X54001, Durban 4000, South Africa }


\begin{abstract}
\noindent  We study the dynamics of test particle and stability of circular geodesics in the gravitational field of a non-commutative geometry inspired Schwarzschild black hole spacetime (NCSBH). The coordinate time Lyapunov exponent ($\lambda_{c}$) is crucial to investigate the stability of equatorial circular geodesics of massive and massless test particles. The stability or instability of circular orbits are discussed by analysing the variation of Lyapunov exponent with radius of these orbits for different values of non-commutative parameter ($\alpha$). In the case of null circular orbits, the instability exponent is calculated and presented to discuss the instability of null circular orbits.  Further, by relating parameters corresponding to null circular geodesics (i.e. angular frequency and Lyapunov exponent), the quasinormal modes (QNMs) for a massless scalar field perturbation in the eikonal approximation are evaluated, and also visualised by relating the real and imaginary parts. The nature of scalar field potential, by varying the non-commutative parameter ($\alpha$) and angular momentum of perturbation ($l$), are also observed and discussed. \\
\end{abstract}
\maketitle
\section{Introduction}
The most celebrated theory of gravity so-called the general relativity theory (GR) has many successful predictions including the fascinating black holes (BHs) which are obtained as unique solutions of Einstein’s field equations \cite{Hartle2003}. Among the various observational phenomenons detected before, the first BH image by Event Horizon Telescope (EHT) also confirms Einstein's theory of GR which actually shows the shadow of the supermassive BH in the center of Messier 87 (M87), an elliptical galaxy about 55 million light years from earth \cite{event2019first}. The study of geodesic motion of test particles in the gravitational field of BH spacetimes is an exciting issue among astrophysicists in view of observing various physical properties of spacetime.  In particular, the curvature of any BH spacetime is analysed through geodesics around it to obtain a rich structure with conveying the characteristics of that particular spacetime geometry \cite{Hartle2003,Wald1984c,poisson2004relativist}.\\
\noindent In last few decades, several studies have been done concerning the non-commutative geometry in GR as an intrinsic property of spacetime which does not depend on curvature of spacetime and gained interest in context of approaching gravity at quantum level \cite{ansoldi2007non,nicolini2006noncommutative,brown2011instability,grezia2011non,rahaman2015particle,rahaman2013btz,kuniyal2018null,alavi2009reissner,sharif2012non,arraut2009noncommutative,chabab2012schwarzschild,ma2017noncommutative}. Following the non-commutative Heisenberg algebra \cite{douglas2001noncommutative}, the non-commutative inspired geometry in GR can be interpreted as an uncertainty in the spatial coordinates defined by the commutation relation $[x^{\mu},x^{\nu}]=i \alpha^{\mu\nu}$, where $ \alpha^{\mu\nu}$ being an antisymmetric matrix \cite{rahaman2013btz,alavi2009reissner,ahluwalia1994quantum,conroy2003phenomenology}. It is a topic of keen interest due to the non-commutativity effect on spacetime as a consequence of which  point-like matter structures are turned into smeared objects diffused throughout a region \cite{rahaman2015particle,kuniyal2018null}.  More precisely, non-commutative geometries have been constructed as an approach to quantum gravity by eliminating the singularities which appear in GR. In view of non-commutativity, the quantum effect appears only in the matter source such that the mass of the particle will not be localized at the point while the Einstein tensor remains the same. The position Dirac delta function of point-like mass density source is eliminated by Gaussian distribution having a minimal width of $\sqrt{\alpha}$ representing smeared objects.  The conventional mass density of a static, spherically symmetric, smeared particle-like gravitational source is defined in form of Gaussian distribution as \cite{rahaman2015particle,ma2017noncommutative}

\begin{equation}
\rho_{\alpha}(r)= \frac{M}{(4\pi \alpha)^{3/2}}e^{-r^{2}/4\alpha},\label{eq1}
\end{equation} 

\noindent where, $\alpha$  is called the commutative parameter representing the non-commutativity of spacetime and has a dimension of length squared.\\
The main objective of our study is to demonstrate how spatial noncommutativity effects the stability of circular orbits and the characteristics modes or QNMs in the gravitational field of NCSBH arises due to non-commutative structure of spacetime. These have interesting features due to the non-commutativity of spacetime. Null circular geodesics play a crucial role in describing the characteristic modes of a BH, popularly known as QNMs, which can be interpreted as non-massive particles trapped in the unstable circular orbits.\\

\noindent Several kinds of BH solutions inspired by a non-commutative geometry in GR have been studied by Nicolini et al. \cite{nicolini2006noncommutative} which provide fascinating information about their behavior and properties. The motion of massive (i.e. timelike) and massless (i.e. null) test particles in the background of non-commutative black holes, especially Schwarzschild BH, was investigated by Rahaman et al. \cite{rahaman2015particle}. The Reissner-Nordström BH inspired by non-commutative geometry was analysed by Alavi \cite{alavi2009reissner}, and Modesto et al. \cite{modesto2010charged} has done the same investigations for the charged rotating non-commutative BHs, viz the Kerr non-commutative BH and Kerr-Newmann non-commutative BH. So far, the solution of the coupled Einstein-Maxwell field equations with non-commutative geometry which describe charged, self-gravitating objects in diverse context, including extremal and non-extremal BHs, was investigated by  Ansoldi et al. \cite{ansoldi2007non}. The geodesic motion of massless test particles i.e. photons in the background of a non-commutative geometry inspired Schwarzschild BH has been studied by Kuniyal et al. \cite{kuniyal2018null}.\\ 
The QNMs of massless scalar field perturbation in a SBH spacetime inspired by non-commutative geometry has been investigated by adopting the third-order Wentzel–Kramers–Brillouin (WKB) approximation approach which indicates the significant effect of the non-commutative parameter in QNM frequencies \cite{liang2018quasinormal}. Also, the spectrum of QNM frequencies of five-dimensional non-commutative BH spacetime for the perturbations of a massive scalar field using the sixth-order WKB approximation has been recently obtained by Grigoris et al. \cite{panotopoulos2020quasinormal}. However, the massless scalar QNMs of the noncommutative D-dimensional Schwarzschild-Tangherlini BH spacetime has also recently been investigated employing the WKB approximation method, the asymptotic iterative method (AIM) and the inverted potential method (IPM) in a greater detail by Zening et al. \cite{yan2020scalar}.\\

\noindent Ćirić et al. \cite{ciric2018noncommutative} have performed a detailed study of QNM spectrum of the Reissner–Nordström BH in the presence of a deformed spacetime structure which is obtained from a noncommutative deformation of a scalar field coupled with a commutative Reissner–Nordström spacetime. Thereafter, they have used a numerical method known as continued fraction method to obtain the QNM spectrum for a non-extremal Reissner–Nordström BH spacetime (see \cite{ciric2019noncommutative} for details) and compared the obtained results with their previous results as in \cite{ciric2018noncommutative}. However, Gupta et al. \cite{gupta2015noncommutative}  have obtained an exact analytic expression for the QNMs of a noncommutative massless scalar field in the gravitational field of 2+1 dimensional BTZ BH spacetime up to the first order in the deformation parameter and quantization of BH entropy \cite{gupta2014effects,liu2009quantization}. Further, the fermionic QNMs of the BTZ BH spacetime in the view of spacetime noncommutativity which leads a duality between a spinless and spinning BTZ BH case is also analysed in detail \cite{gupta2017noncommutative}. More recently, the noncommutativity and quantum corrections to the Hawking temperature and entropy of a BTZ BH spacetime are investigated in \cite{anacleto2021noncommutative} by applying Hamilton-Jacobi method using the WKB approximation which indicates that the logarithmic correction to the entropy of the noncommutative BTZ BH is direct evidence of noncommutativity of spacetime.\\
Among the various methods and techniques for analysing the stability of geodesics, the measurement of Lyapunov exponent has been extensively employed. The average rate of separation between two nearby trajectories in phase space is defined as Lyapunov exponent. The Lyapunov exponent should have a positive and negative value corresponding to divergence and convergence of two neighbouring geodesics respectively \cite{cardoso2009geodesic,sharif2017particle,pradhan2016stability,mondal2020geodesic,pradhan2015stability}. \\
A simple set of unstable circular orbits exist in the extreme case of BH in GR and other alternative theories of gravity including the usual stable circular orbits. This set of unstable circular orbits in any gravitational field of BH is a direct consequence of the non-linearity of GR which is quantified by a positive value of Lyapunov exponent. The nonlinearity in GR leads to nonintegrability of the system and chaos may develop in the unstable circular orbits \cite{cornish2003lyapunov,cornish2001chaos,hilborn2000chaos,Suzuki1997}.
However, the QNM is a complex quantity corresponding to unstable null circular geodesics which can be evaluated by its correlation with Lyapunov exponent and angular frequency \cite{berti2009quasinormal,kokkotas1999quasi,nollert1999quasinormal}. According to Cardoso et al., the real part of the QNM is interpreted as the angular frequency  while the imaginary part is interpreted as the instability timescale of the orbit (i.e. Lyapunov exponent) of the unstable null circular geodesics \cite{cardoso2009geodesic}.\\
This paper is organized as follows. We first discuss the Lyapunov exponent and critical exponent in Section I-(A). The NCSBH spacetime geometry is then briefly discussed in Section II, followed by the stability of geodesics of test particles around considered spacetime in Section III. Further, the massless scalar field perturbation around NCSBH spacetime is discussed along with the computation of QNMs frequencies in Section IV.  Finally, the results obtained are  summarized and concluded in Section V. Throughout the course of this work, we have used rescaled units so that the gravitational constant and the speed of light are normalized (i. e. $ G = c = 1$). The BH mass is however considered as $M=1$ and the non-commutative parameter $\alpha$ is chosen to be in range $0.1$ to $0.3$ while depicting the plots.
\subsection{The Lyapunov exponent and critical exponent}
The concept of Lyapunov exponent has regularly been implemented to study the behavior of dynamical systems especially to determine the unpredictable behavior of non-linear dynamical systems. Basically, the Lyapunov exponent or Lyapunov characteristic exponent characterizes quantitatively the rate of separation of infinitesimally nearby trajectories or orbits say $x(t)$ and $x_{0}(t)$ in phase-space of a dynamical system (see refs.\cite{sandri1996numerical,guan2014important} for details).  \\  
One can thus define an observed trajectory or orbit $x(t)$ representing the solution of a continuous-time smooth dynamical system in any arbitrary dimension with the differential equation  \cite{sano1985measurement},
\begin{equation}
\frac{dx}{dt}=F(x).\label{eq2}
\end{equation}
So far, for a small perturbation $\xi(t)$ on $x(t)$ defined by
\begin{equation}
x(t)= x_{0}+\xi(t),\label{eq3}
\end{equation}
where, $x_{0}$ is a fixed point at $t=0$.

\noindent According to Sano et al. \cite{sano1985measurement}, one can define the principal Lyapunov exponent as the mean exponential rate of expansion or contraction of the trajectory $x_{0}$ in the direction of $\xi(0)$  as
\begin{equation}
\lambda= \lim_{t\to\infty}\frac{1}{t} \text{ln} \frac{\parallel\xi(t)\parallel}{\parallel \xi(0)\parallel},\label{eq7}
\end{equation}
here  $\parallel..\parallel $ represents a vector norm. \\

\noindent Starting from the necessary Lagrangian for the motion of a test particle around the BH spacetime, Cardoso et al. \cite{cardoso2009geodesic} derived a relationship between the second order derivative of the radial effective potential ($\dot{r}^{2}$) and the Lyapunov exponent ($\lambda$).
For the motion of a non-spinning test particle in the gravitational field of any static, spherically symmetric spacetime, the proper time Lyapunov exponent is defined as (see Appendix-A in \cite{giri2021stability} for details)
\begin{equation}
\lambda_{p}=\pm\sqrt{\frac{\left(\dot{r}^{2}\right)^{''}}{2}},\label{eq19}
\end{equation}
and the coordinate time Lyapunov exponent is derived as
\begin{equation}
\lambda_{c}=\pm\sqrt{\frac{\left(\dot{r}^{2}\right)^{''}}{2 \dot{t}^{2}}}.\label{eq20}
\end{equation}
Here and throughout the work $(')$ and $(.)$ represents differentiation w.r.t. radial coordinate $r$ and affine parameter $\tau$ respectively.\\
The unstable circular orbits can be found for the real values of Lyapunov exponents while the stable circular orbits may be found for imaginary values and the marginally stable orbits are found for zero values of the Lyapunov exponent \cite{cardoso2009geodesic}.
Moreover, the critical exponent ($\gamma$) is used to obtain the quantitative characterization of instability of circular orbits defined as the ratio of orbital timescale $T_{\Omega}=\frac{2\pi}{\Omega}$ to the 
Lyapunov timescale $T_{\lambda}=\frac{1}{\lambda}$ in the following form\cite{pretorius2007black,cardoso2009geodesic},
\begin{equation}
\gamma= \frac{T_{\lambda}}{T_{\Omega}}= \frac{\Omega}{2\pi \lambda}\label{eq21}~.
\end{equation}
where $\Omega$ represents angular frequency or orbital angular velocity.

\section{The non-commutative Schwarzschild black hole (NCSBH) } 
The non-commutative geometry is an crucial approach to unify other fundamental forces with gravity revealing the quantum nature of gravity at very high energy regime. The non-commutative inspired BH spacetimes based on the mathematical framework of quantum gravity are one of the leading candidates of this theory \cite{ansoldi2007non,nicolini2006noncommutative}. Considering a matter source having mass density described by Eq.(\ref{eq1}) and solving the Einstein field equation, one can obtained SBH inspired by non-commutative geometry (i.e. NCSBH spacetime) defined by a line element in standard spherical coordinates ($t,r,\theta,\phi$) given as \cite{ansoldi2007non,rahaman2015particle,kuniyal2018null,rahaman2013btz,ma2017noncommutative}

\begin{equation}
ds^{2}= -f(r) dt^{2}+\frac{d r^{2}}{f(r)} + r^{2} \left(d\theta^{2}+\sin^{2}\theta d\phi^{2}\right)~,\label{eq23}
\end{equation}
where, the  metric function $f(r)$ reads as\cite{rahaman2015particle,alavi2009reissner},
\begin{equation}
f(r)= 1-\frac{2 m(r)}{r} \label{metricfn}.
\end{equation}

\noindent The mass function $m(r)$, which is proportional to the mass $ M $ of BH, and related to the non-commutative parameter $\alpha$, by a relation
\begin{equation}
m(r)= \frac{2M}{\sqrt{\pi}} ~\gamma (\frac{3}{2},\frac{r^{2}}{4\alpha}).
\end{equation}
Here, the lower incomplete gamma function is given by \cite{jameson2016incomplete}
\begin{equation}
\gamma (\frac{3}{2},\frac{r^{2}}{4\alpha})= \int_{0}^{\frac{r^{2}}{4\alpha}}\sqrt{t} e^{-t}dt.
\end{equation}

\noindent The lower and upper incomplete gamma functions satisfy the following relation

\begin{equation}
\gamma (\frac{3}{2},\frac{r^{2}}{4\alpha})+\Gamma (\frac{3}{2},\frac{r^{2}}{4\alpha})=\frac{\sqrt{\pi}}{2} .
\end{equation}

\noindent Hence, the metric function Eq.(\ref{metricfn}) can be rewritten in terms of the upper incomplete gamma function as
\begin{equation}
f(r)= 1-\frac{2 M}{r}+\frac{4M}{\sqrt{\pi} r} \Gamma (\frac{3}{2},\frac{r^{2}}{4\alpha}),
\end{equation}
i.e. the third term represents the perturbation due to non-commutativity of spacetime.
\begin{figure}[H]
	\centering
	\includegraphics[width=09cm,height=8cm]{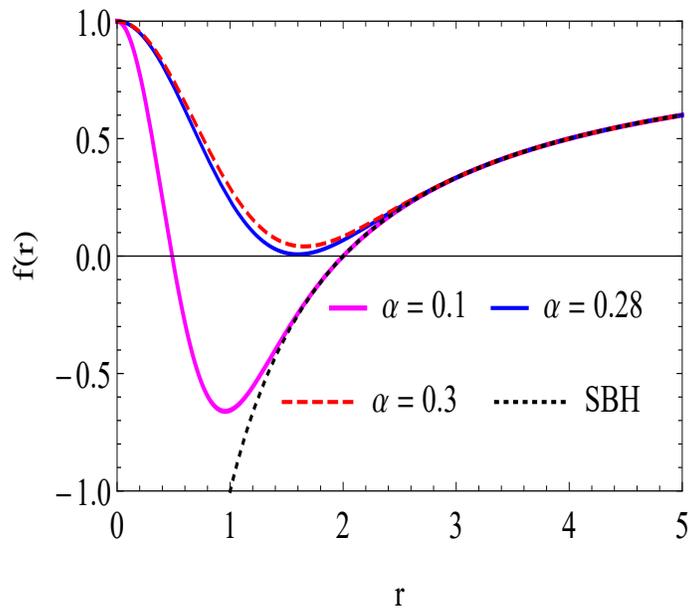}
	\caption { The variation of  $``f(r)"$ as a function of $``r"$ for different values of non-commutative parameter $\alpha$ with $M=1$ and for SBH in limit $\alpha \to 0$.   }\label{HorizonNCSBH}
\end{figure}
\noindent However, the SBH spacetime can be recovered from NCSBH metric in the prescribed limit $\alpha \to 0$ or $r/\sqrt{\alpha} \to \infty$ as the gamma function $\Gamma (\frac{3}{2},\frac{r^{2}}{4\alpha})$ vanishes in this limit.

\noindent The position of the horizon radius ($r_{h}$) at which the metric function $f(r)$ vanishes, is presented in \figurename{\ref{HorizonNCSBH}}. One can conclude that the non-commutative inspired SBH spacetime introduces two horizons for $\alpha=0.1$, one horizon for $\alpha=0.28$, no horizon for $\alpha=0.3$. Contrary to NCSBH,  the SBH has only one horizon at $r=2M$. 
\section{Geodesic stability around NCSBH spacetime}
\noindent In this section, we first investigate the affinely parametrized geodesics (timelike and null) by the Lagrangian approach i.e. the Euler-Lagrange equations \cite{Hartle2003,poisson2004relativist}. For this purpose, we begin with the Lagrangian defined as below
\begin{equation}
\mathcal{L} =\frac{1}{2} g_{\mu\nu}\dot{x^{\mu}}\dot{x^{\nu}},
\end{equation}
where, $g_{\mu\nu}$ represents the metric tensor and $\dot{x^{\mu}}$ refers derivative of spacetime coordinates w.r.t. affine parameter $\tau$. 
Therefore, the Lagrangian for test particles motion in NCSBH background considering the motion in equatorial plane (i.e. $\theta=\frac{\pi}{2}$) can be written as
	\begin{equation}
	2\mathcal{L}= - \left[1-\frac{2 M}{r}+\frac{4M}{\sqrt{\pi} r} \Gamma (\frac{3}{2},\frac{r^{2}}{4\alpha})\right]
	\dot{t}^{2}+\frac{\dot{r}^{2}}{\left[ 1-\frac{2 M}{r}+\frac{4M}{\sqrt{\pi} r} \Gamma (\frac{3}{2},\frac{r^{2}}{4\alpha})\right]} + r^{2} \dot{\phi}^{2}. \label{Lagrangian}
	\end{equation}

\noindent Corresponding to the cyclic coordinates $t$ and $\phi$ in the above Lagrangian, the constants of motion $E$ and angular momentum $L$ are identified. Using the Euler-Lagrange equations of motion, the canonical momenta with respect to these cyclic coordinates are deduced as
\begin{equation}
p_{t}=-\left[1-\frac{2 M}{r}+\frac{4M}{\sqrt{\pi} r} \Gamma (\frac{3}{2},\frac{r^{2}}{4\alpha})\right] \dot{t}
=-E ,\label{pt}
\end{equation}

\begin{equation}
p_{\phi}=r^{2}\dot{\phi}=L .\label{pphi}
\end{equation}
From the above two equations Eqs.(\ref{pt}) and (\ref{pphi}), the first integrals of the geodesic equations yield
\begin{equation}
\dot{t}=\frac{E}{\left[1-\frac{2 M}{r}+\frac{4M}{\sqrt{\pi} r} \Gamma (\frac{3}{2},\frac{r^{2}}{4\alpha})\right]} ,~~~~~
\dot{\phi}= \frac{L}{r^{2}}.\label{firstinte}
\end{equation}

\noindent The constraint on the geodesics reads as
\begin{equation}
g_{\mu \nu} \dot{x}^{\mu}\dot{x}^{\nu}=\epsilon ,
\end{equation}
i.e.

	\begin{equation}
	- \left[1-\frac{2 M}{r}+\frac{4M}{\sqrt{\pi} r} \Gamma (\frac{3}{2},\frac{r^{2}}{4\alpha})\right]
	\dot{t}^{2}+\frac{\dot{r}^{2}}{\left[ 1-\frac{2 M}{r}+\frac{4M}{\sqrt{\pi} r} \Gamma (\frac{3}{2},\frac{r^{2}}{4\alpha})\right]} + r^{2} \dot{\phi}^{2}=\epsilon, \label{constraint}
	\end{equation}

\noindent where, $\epsilon$ = 0 and -1  corresponds to null (massless particles) and timelike (massive particles) geodesics respectively.

\noindent The radial geodesic equation of test particles for NCSBH can be obtained by substituting $\dot{t}$ and $\dot{\phi}$ from Eq.(\ref{firstinte}) into constraint Eq.(\ref{constraint}) as follows

\begin{equation}
\dot{r}^{2}= E^{2} -\left[1-\frac{2 M}{r}+\frac{4M}{\sqrt{\pi} r} \Gamma (\frac{3}{2},\frac{r^{2}}{4\alpha})\right]
\left(\frac{L^{2}}{r^{2}}-\epsilon\right).\label{radialeq}
\end{equation}
Now, by comparing the above Eq.\eqref{radialeq} with $\dot{r}^{2}=E^{2}- V_{eff}$, we found 

\begin{equation}
V_{eff}= \left[1-\frac{2 M}{r}+\frac{4M}{\sqrt{\pi} r} \Gamma (\frac{3}{2},\frac{r^{2}}{4\alpha})\right]
\left(\frac{L^{2}}{r^{2}}-\epsilon\right),\label{}
\end{equation}
which is called the effective potential for the NCSBH spacetime.
\subsection{ The timelike geodesics and Lyapunov exponent }
From Eq.(\ref{radialeq}), when $\epsilon=-1$ taken into account, the radial equation for massive test particles becomes
\begin{equation}
\dot{r}^{2}= E^{2} -\left[1-\frac{2 M}{r}+\frac{4M}{\sqrt{\pi} r} \Gamma (\frac{3}{2},\frac{r^{2}}{4\alpha})\right]
\left(1+\frac{L^{2}}{r^{2}}\right).\label{radialtime}
\end{equation}
Thus, the effective potential of NCSBH for massive (timelike) test particles reads as
\begin{equation}
V_{eff}^{time}= \left[1-\frac{2 M}{r}+\frac{4M}{\sqrt{\pi} r} \Gamma (\frac{3}{2},\frac{r^{2}}{4\alpha})\right]
\left(1+\frac{L^{2}}{r^{2}}\right).\label{effectivetime}
\end{equation}

\begin{figure}[H]
	\centering
	\subfigure[]{\includegraphics[width=08cm,height=7.5cm]{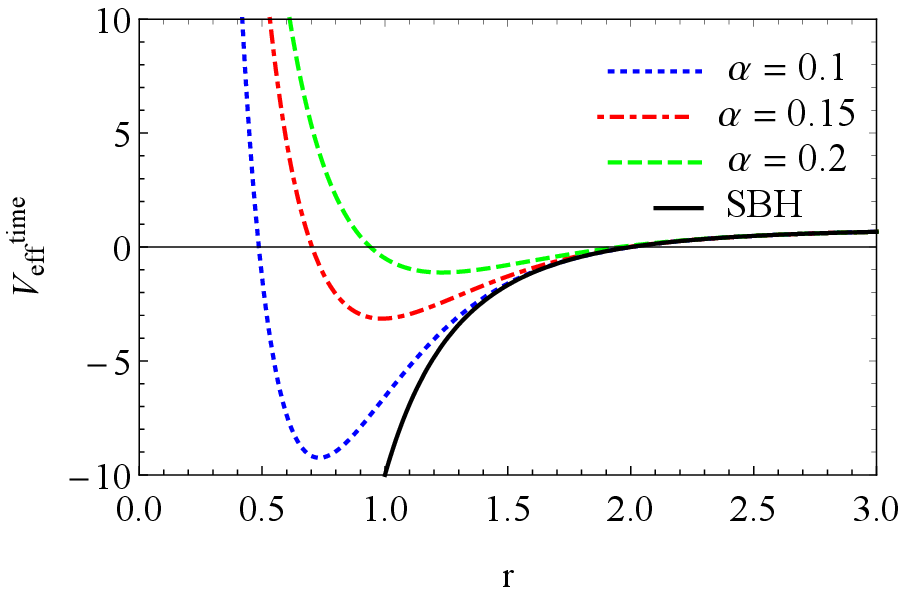}}
	\subfigure[]{\includegraphics[width=08cm,height=7.5cm]{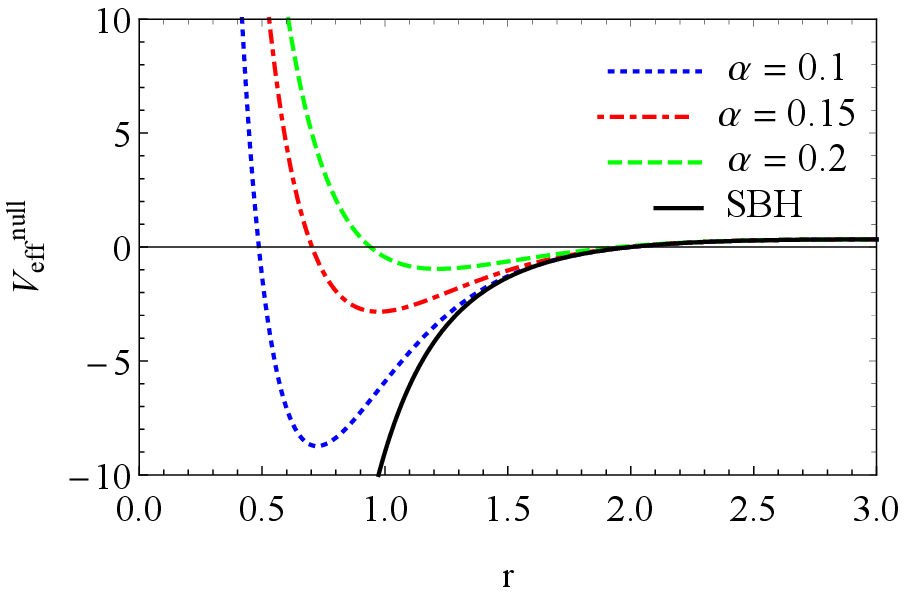}}
	\caption { The behavior of  $``V_{eff}"$  with respect to $``r"$ for massive (see left panel) and massless (see right panel) particles with different values of parameters $\alpha$ and fixed mass $M=1$. The same is also presented for SBH at $\alpha \to 0$. \label{effpot} }
\end{figure}

\noindent As shown in \figurename{\ref{effpot}}, the minima of the effective potential shifted towards higher values of radius as non-commutative parameter $\alpha$ increases. It is also observed that potentials coincide with potential of SBH for radius greater than the horizon of SBH (i.e. $\geq2M$). Thus it can be concluded that the effect of non-commutativity is insignificant for radius  $r\geq2M$. \\
Further, we consider circular geodesic motion of particles in specified NCSBH background by restricting radius of orbits to $r = r_{0}= constant$.\\
The condition for occurrence of circular  circular orbits ( i.e. $\dot{r}^{2}=(\dot{r}^{2})'=0$) and the radial equation expressed in Eq. (\ref{radialtime})  results in the energy and angular momentum describing a circular orbit of the massive particle respectively, as given below,

\begin{equation}
E_{0}^{2}= f(r_{0})  \left(1 - \Psi \right),\label{energy}
\end{equation}

\begin{equation}
L_{0}^{2}= - \Psi   r_{0}^{2}\label{angmomentum},
\end{equation}

\noindent where the term $\Psi$ associated with the above equations is represented as

	\begin{equation}
	\Psi= \frac{M \left(r_{0}^{2} \sqrt{\frac{r_{0}^{2}}{\alpha}} - 
		2 e^{\frac{r_{0}^{2}}{4\alpha}} \sqrt{\pi} \alpha + 4 e^{\frac{r_{0}^{2}}{4\alpha}} \alpha \Gamma \left(\frac{3}{2},\frac{r_{0}^{2}}{4\alpha}\right)\right) }{M r_{0}^{2} \sqrt{\frac{r_{0}^{2}}{\alpha}} - 6  e^{\frac{r_{0}^{2}}{4\alpha}} M \sqrt{\pi} \alpha + 2 e^{\frac{r_{0}^{2}}{4\alpha}} \sqrt{\pi} r_{0} \alpha + 12 e^{\frac{r_{0}^{2}}{4\alpha}} M \alpha \Gamma (\frac{3}{2},\frac{r_{0}^{2}}{4\alpha})}.\label{Psi}
	\end{equation}

\noindent The circular orbits are possible for real and finite values of the energy and angular momentum, for which the quantity $\Psi$ must be negative i.e. $\Psi <0$.\\
However, the angular frequency ($\Omega_{0}$) for timelike circular orbits is deduced as

\begin{equation}
\Omega_{0} = \frac{\dot{\phi}}{\dot{t}} = \frac{f(r_{0})}{r_{0}^{2}}\frac{L_{0}}{E_{0}}  .\label{angufreq}
\end{equation}

\noindent In order to discuss the stability of circular orbits of massive particles around NCSBH,  by calculating the second order derivative of the Eq.(\ref{radialtime}) with respect to radius ($r$), the coordinate time Lyapunov exponent represented by Eq.(\ref{eq20}), is evaluated as

	\begin{equation}
	\lambda_{c} =\sqrt{\frac{ (-1+\Psi)f(r_{0})\left[\frac{e^{\frac{r_{0}^{2}}{4\alpha} M r_{0} \sqrt{\frac{r_{0}^{2}}{\alpha}}} }{2 \sqrt{\pi} \alpha^{2}} -\frac{8M \left(\frac{\sqrt{\pi}}{2}-\Gamma \left(\frac{3}{2},\frac{r_{0}^{2}}{4\alpha}\right)\right) }{\sqrt{\pi}r_{0}^{3}}\right]-\frac{4\Psi}{r_{0}}f(r_{0})\left[-\frac{e^{\frac{r_{0}^{2}}{4\alpha}}M \sqrt{\frac{r_{0}^{2}}{\alpha}} }{\sqrt{\pi}\alpha}+\frac{4M \left(\frac{\sqrt{\pi}}{2}-\Gamma \left(\frac{3}{2},\frac{r_{0}^{2}}{4\alpha}\right)\right) }{\sqrt{\pi}r_{0}^{2}}\right]+ \frac{6\Psi f(r_{0})^{2}}{r_{0}^{2}}}{2(1-\Psi)}} .\label{eq37}
	\end{equation}\label{LyExtime}

\begin{figure}[H]
	\centering
	\includegraphics[width=9cm,height=8cm]{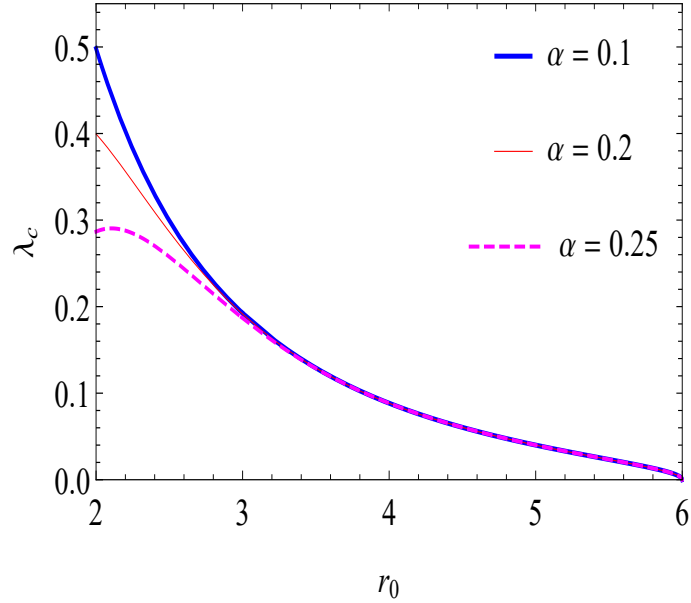}
	\caption { The Lyapunov exponent ``$\lambda_{c}$" as a function of radius of orbit ``$r_{0}$" for different values of parameter $\alpha$ (labeled in figure) for mass $M=1$. }\label{LambdaCtime}
\end{figure}

\noindent One can illustrate the stability or instability of the timelike circular orbits for NCSBH by Lyapunov exponent computed above. The orbits are stable for complex nature (i.e. imaginary) of $\lambda_{c}$, and unstable orbits exist for real value of $\lambda_{c}$. The nature of Lyapunov exponent $\lambda_{c}$ with radius of circular orbits by varying non-commutative parameter $\alpha$ is depicted in \figurename{\ref{LambdaCtime}}, which indicates the instability defined by $\lambda_{c}$ decreases with increase in radius of circular orbits and coincide at higher values of radius.

\subsection{ The null geodesics and Lyapunov exponent }
Substitute $\epsilon=0$ into Eq.(\ref{radialeq})	in order to investigate null circular geodesics for NCSBH.  The radial equation for photons (massless particles) reads as

\begin{equation}
\dot{r}^{2}=E^{2} -\left[1-\frac{2 M}{r}+\frac{4M}{\sqrt{\pi} r} \Gamma (\frac{3}{2},\frac{r^{2}}{4\alpha})\right]
\left(\frac{L^{2}}{r^{2}}\right).\label{radialnull}
\end{equation}
The effective potential for photons is then obtained as
\begin{equation}
V_{eff}^{null}=\left[1-\frac{2 M}{r}+\frac{4M}{\sqrt{\pi} r} \Gamma (\frac{3}{2},\frac{r^{2}}{4\alpha})\right]
\left(\frac{L^{2}}{r^{2}}\right),\label{nulleffpot}
\end{equation}
and the same is depicted in \figurename{\ref{effpot}} with same observations as that for timelike case.

\noindent The null circular orbits are possible when $\dot{r}^{2}=0$ at constant radius $r=r_{c}$ which yield the angular momentum to energy ratio (i.e. impact parameter) as follows 
\begin{equation}
D_{c}= \frac{L_{c}}{E_{c}}= \sqrt{\frac{r_{c}^{2}}{f(r_{c})}},
\end{equation}

\noindent and  $(\dot{r}^{2})'=0$  leads to an equation for the radius of unstable circular orbits as
\begin{equation}
r_{c}^{2} e^{-\frac{r_{c}^{2}}{4\alpha}}-\frac{12 \alpha^{3/2}}{r_{c}}\left[\frac{\sqrt{\pi}}{2}-\Gamma (\frac{3}{2},\frac{r_{c}^{2}}{4\alpha})\right]+2\frac{\alpha^{3/2}\sqrt{\pi}}{M}=0.\label{eq45}
\end{equation}

\noindent The solution of the above equation provides the radius of the unstable null circular orbit or photon sphere given as

\begin{equation}
r_{c}= 3M \left(1-\frac{M}{\sqrt{\pi}\alpha}e^{-\frac{M^{2}}{\alpha}}\right) .\label{eq46}
\end{equation}

\noindent Thus the radius of unstable circular orbit of photons for NCSBH is smaller than that for SBH (i.e. $<3M$) due to the presence of non-commutative geometry in spacetime.  Also, both the impact parameter ($D_{c}$) and $r_{c}$ of null circular orbits for NCSBH are dependent on mass and non-commutative parameter.

\noindent The angular frequency at $r=r_{c}$ comes out as
\begin{equation}
\Omega_{c}= \frac{\dot{\phi}}{\dot{t}}= \sqrt{\frac{f(r_{c})}{r_{c}^{2}}}=\frac{1}{D_{c}} .\label{eq47}
\end{equation}
\noindent So, it is interesting to say that the angular frequency of null circular geodesics is equivalent to inverse of the impact parameter.\\
Further, we derived the coordinate time Lyapunov exponent for null circular orbits by using Eqs.(\ref{eq20}) and (\ref{radialnull}) as given below

	\begin{equation}
	\lambda_{Null}= \sqrt{\frac{f(r_{c})}{2  r_{c}^{2}}\left[\frac{24 M}{\sqrt{\pi}r_{c}}\left(\frac{\sqrt{\pi}}{2}-\Gamma \left(\frac{3}{2},\frac{r_{c}^{2}}{4\alpha}\right)\right)- \frac{e^{-\frac{r_{c}^{2}}{4\alpha}}M r_{c}^{3} \sqrt{\frac{r_{c}^{2}}{\alpha}}}{2\sqrt{\pi}\alpha^{2}}-4 \frac{e^{-\frac{r_{c}^{2}}{4\alpha}}M r_{c} \sqrt{\frac{r_{c}^{2}}{\alpha}}}{\sqrt{\pi}\alpha} -6 f(r_{c})\right]}.\label{LYEXnull}
	\end{equation}
	
\noindent The null circular orbits are unstable at the radius $r_{c}$ for real values of Lyapunov exponent $\lambda_{Null}$. The instability of unstable null circular orbits can be determined by a quantity known as instability exponent which is defined as the ratio of Lyapunov exponent to angular frequency ($ \lambda_{Null}/\Omega_{c}$) and comes out as follows
	\begin{equation}
	\frac{\lambda_{Null}}{\Omega_{c}}= \sqrt{\frac{1}{2}\left[\frac{24 M}{\sqrt{\pi}r_{c}}\left(\frac{\sqrt{\pi}}{2}-\Gamma \left(\frac{3}{2},\frac{r_{c}^{2}}{4\alpha}\right)\right)- \frac{e^{-\frac{r_{c}^{2}}{4\alpha}}M r_{c}^{3} \sqrt{\frac{r_{c}^{2}}{\alpha}}}{2\sqrt{\pi}\alpha^{2}}-4 \frac{e^{-\frac{r_{c}^{2}}{4\alpha}}M r_{c} \sqrt{\frac{r_{c}^{2}}{\alpha}}}{\sqrt{\pi}\alpha} -6 f(r_{c})\right]}.
	\end{equation}\label{InstExp}

\begin{figure}[H]
\centering
\subfigure[]{\includegraphics[width=8cm,height=7.0cm]{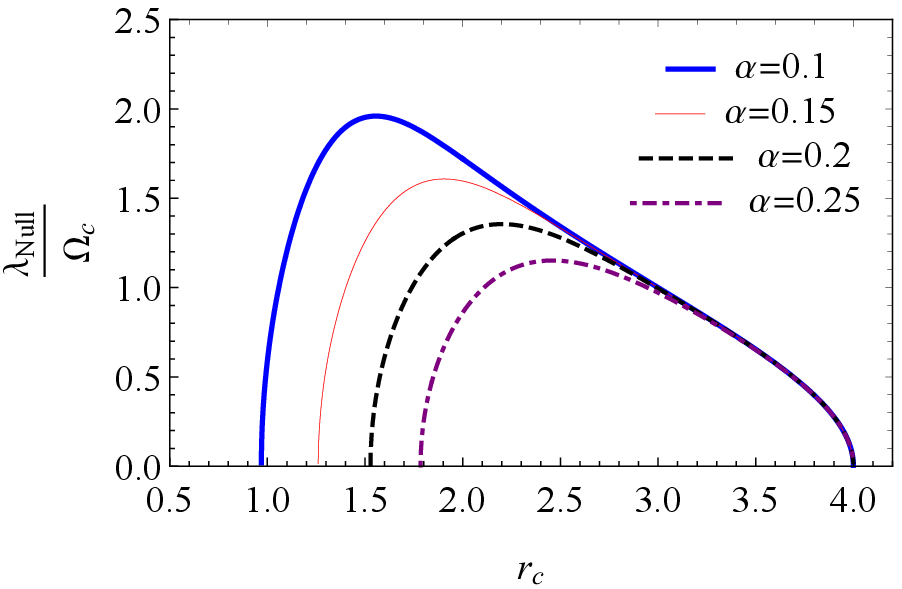}\label{InstExponent}}
\subfigure[]{\includegraphics[width=8cm,height=7.0cm]{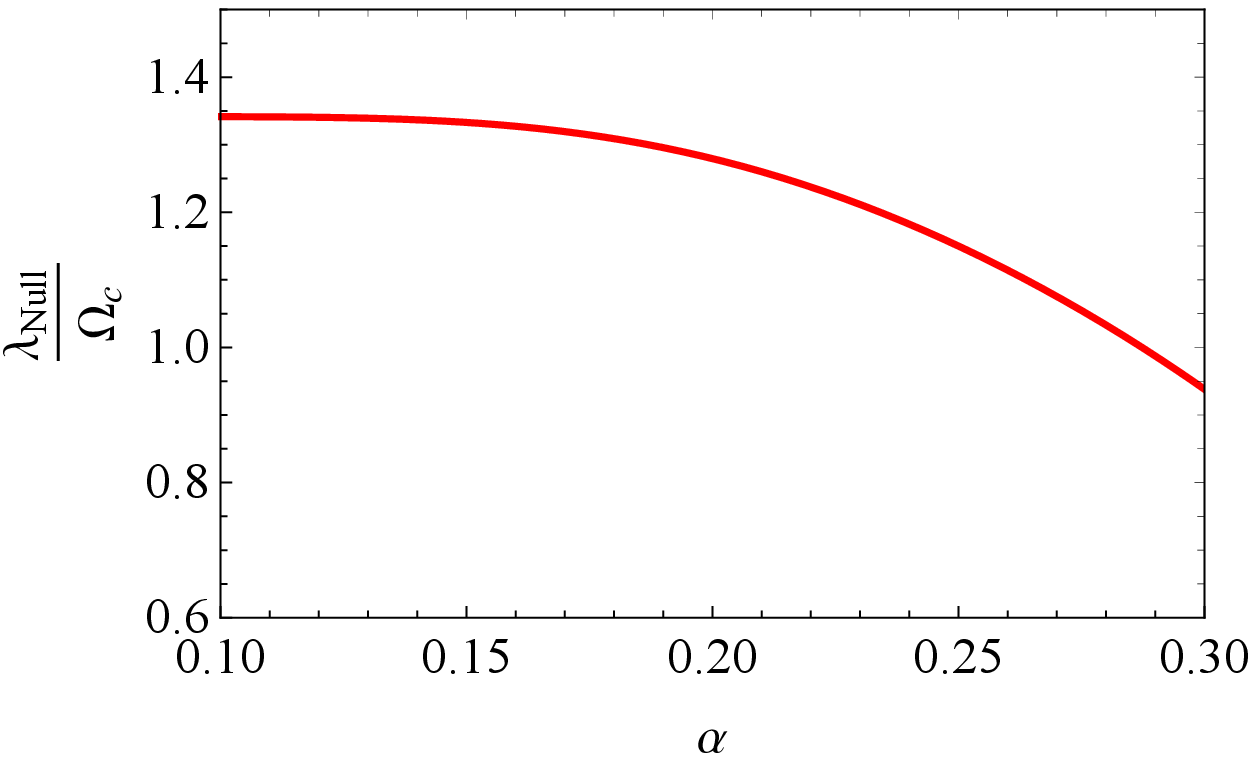}\label{InstExponentAlph}}	\caption{(a) The variation of the instability exponent ``$(\frac{\lambda_{Null}}{\Omega_{c}})$" with radius ``$r_{c}$" for various values of parameter $\alpha$ (labeled in figure) and fixed mass $M=1$. (b) ``$(\frac{\lambda_{Null}}{\Omega_{c}})$" as a function of  $\alpha$ for a circular orbit of radius $r_{c}=2.5$ and $M=1$. }
\end{figure}

\noindent The \figurename{\ref{InstExponent}} depicts the variation of instability exponent for null circular orbits with respect to radius $r_{c}$. For various values of non-commutative parameter $\alpha$, it can be observed that instability of increases sharply to highest value then decreases as radius of circular orbit increases. Also, the height of instability decreases as value of the parameter $\alpha$ increases and shows the same behaviour for values of radius ($r_{c}>3M$). Therefore, one can state that the null circular orbits in NCSBH spacetime are found unstable in nature. Furthermore, in \figurename{\ref{InstExponentAlph}}, we visualize the variation of instability of circular orbit of radius $r_{c}=2.5$ with respect to $\alpha$. It is observed that the instability of circular orbit first remain constant for smaller values of $\alpha$ and then decreases gradually as values of $\alpha$ increases.
\noindent Further, we deduced the critical exponent that measures the instability of null circular orbits quantitatively as
	\begin{equation}
	\gamma_{Null} =\frac{\Omega_{c}}{2\pi \lambda_{Null}} =\frac{1}{2\pi \sqrt{\frac{1}{2}\left[\frac{24 M}{\sqrt{\pi}r_{c}}\left(\frac{\sqrt{\pi}}{2}-\Gamma \left(\frac{3}{2},\frac{r_{c}^{2}}{4\alpha}\right)\right)- \frac{e^{-\frac{r_{c}^{2}}{4\alpha}}M r_{c}^{3} \sqrt{\frac{r_{c}^{2}}{\alpha}}}{2\sqrt{\pi}\alpha^{2}}-4 \frac{e^{-\frac{r_{c}^{2}}{4\alpha}}M r_{c} \sqrt{\frac{r_{c}^{2}}{\alpha}}}{\sqrt{\pi}\alpha} -6 f(r_{c})\right]}} .\label{eq50}
	\end{equation}

\noindent In the case of null geodesics, the Lyapunov exponent $\lambda_{Null}^{Ele}$ has a real value at $r=r_{c}$. Therefore, the Lyapunov timescale is less than orbital timescale ($T_{\lambda}<T_{\Omega}$) of orbits which implies the  observational appearance of instability in the null circular orbits for NCSBH. 

\section{Massless scalar field perturbation around NCSBH spacetime}
In this section, we analyze the behavior of  NCSBH geometry  in view of perturbations by a massless scalar field following S. Fernando \cite{fernando2017bardeen,fernando2015regular,fernando2015quasi}. The massless scalar field perturbation in eikonal approximation has been widely used to evaluate QNMs of various spacetime background due to its considerable accuracy. \\
The Klein-Gordon (K-G) equation describing the motion of a test particle in scalar field around any BH has the form
\begin{equation}
(\Delta^{2}-\mu^{2})\chi=0. \label{KGeq}
\end{equation}
As we consider massless scalar field perturbations so that $\mu=0$, the K-G equation reduces to
\begin{equation}
\Delta^{2}\chi=0. \label{KGeq2}
\end{equation}
The above K-G equation (\ref{KGeq2}) leads to a scalar field with decomposition 
\begin{equation}
\chi= e^{-i\omega t} Y_{l,m}(\theta,\phi)\frac{R(r)}{r},
\end{equation}

\noindent where, $Y_{l,m}(\theta,\phi)$ represents the spherical harmonics, $l $ is angular quantum number and $m$ the magnetic quantum number, whereas the frequency of the oscillations of the scalar field is represented by $\omega$.\\
The radial component of the Eq.(\ref{KGeq}) can be simplified to a Schrödinger-type equation as below
\begin{equation}
\frac{d^{2}R(r_{*})}{dr^{2}_{*}}+\left(\omega^{2}-V_{s}(r)\right)R(r_{*})=0,
\end{equation}
where,  $r_{*}$ is “tortoise” coordinate ranging from $-\infty$ to $+\infty$ and is defined as $dr_{*}=\frac{dr}{f(r)}$.	\\
The function $V_{s}(r)$ is known as the scalar field potential for massless scalar field perturbation having the angular momentum of the perturbation $l$. In view of null circular orbit at $r=r_{c}$ for eikonal limit $(l \to  \infty)$, it is obtained as
\begin{equation}
V_{s}(r_{c})\approx l \left(\frac{E_{c}}{L_{c}}\right)^{2}=l \Omega_{c}^{2}.\label{eq76}
\end{equation}

\noindent Hence, the scalar field potential in case of NCSBH geometry is given by
\begin{equation}
V_{s}(r_{c}) = l \frac{f(r_{c})}{r_{c}^{2}}= \frac{l}{r_{c}^{2}}\left( 1-\frac{2 M}{r_{c}}+\frac{4M}{\sqrt{\pi} r_{c}} \Gamma (\frac{3}{2},\frac{r_{c}^{2}}{4\alpha})\right).\label{sclrpot}
\end{equation}
\begin{figure}[H]
\centering
	\subfigure[]{\includegraphics[width=8cm,height=7.5cm]{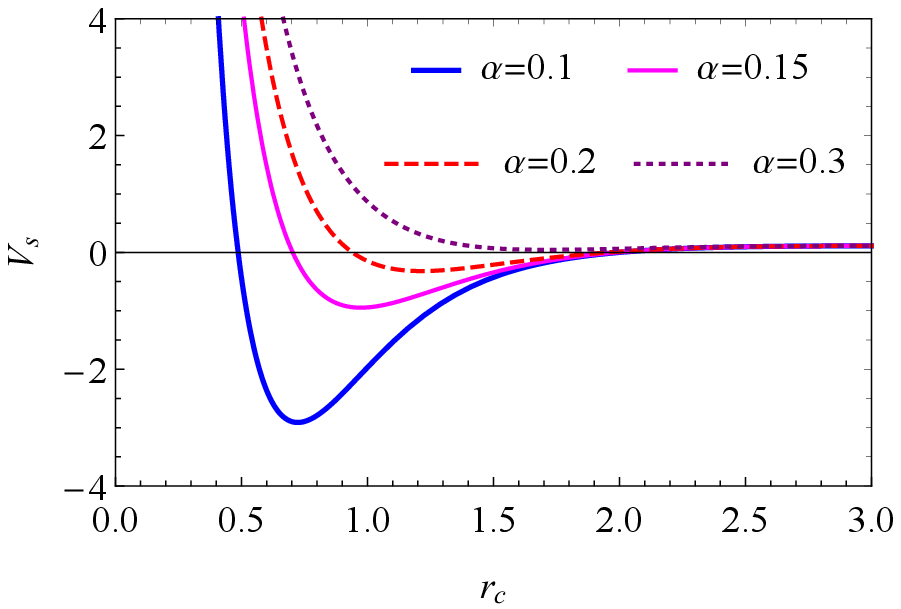}\label{Vs1}}
	\subfigure[]{\includegraphics[width=8cm,height=7.5cm]{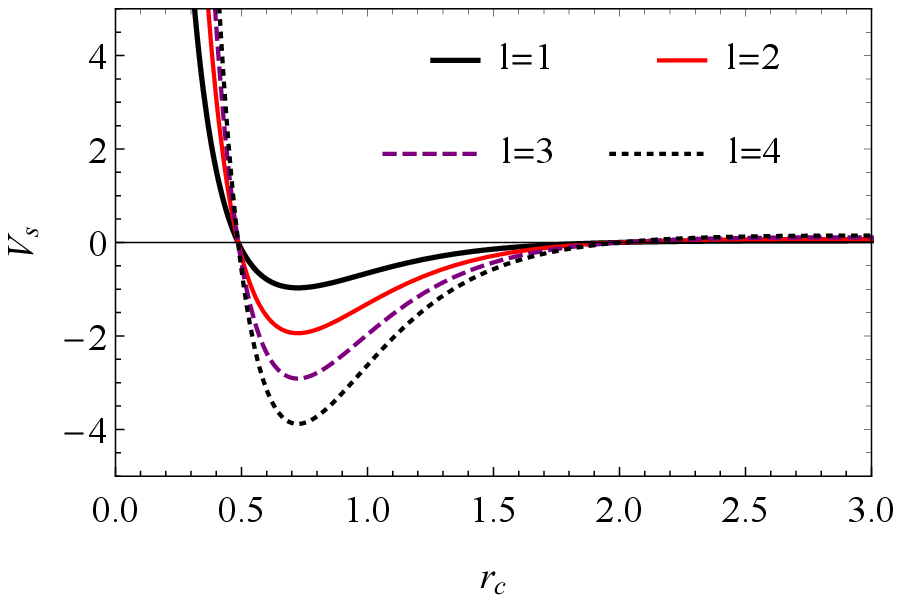}\label{Vs2}}
	\caption{The scalar field potential $``V_{s}"$ versus $``r_{c}"$ by varying non-commutative parameter $\alpha$ for fixed $l=3$ and $M=1$ (see left panel), and by varying perturbation parameter $l$ for fixed values of $\alpha=0.1$ and $M=1$ (see right panel).}
\end{figure}

\noindent Thus, one can notice that the scalar field potential for this geometry depends on the mass of BH ($M$), non-commutative parameter ($\alpha$) and angular momentum of the perturbation ($l$). \figurename{\ref{Vs1}} shows the effect of parameter ($\alpha$) on the nature of the scalar potential for fixed values of perturbation parameter ($l=3$) and BH mass $M=1$. It is observed that the potential decreases to minimum value then increases to positive region whereas the depth of the potential increases as the value of $\alpha$ decreases. However, the depth of scalar potential increases with an increase in angular momentum of perturbation $l$ as shown in \figurename{\ref{Vs2}} for fixed values of parameters $M=1$ and $\alpha=0.1$. Consequently, the negative scalar field potential indicates the presence of unstable null circular orbits around NCSBH spacetime.
\subsection{QNMs of NCSBH Spacetime}
According to Cardoso et al. \cite{cardoso2009geodesic}, the QNMs or  “free” modes of vibration of any static, spherically symmetric, asymptotically flat BH spacetime can be interpreted by unstable null circular orbits corresponding to that spacetime background. In view of the eikonal approximation  i.e. for large limit ($l>>1$), the QNM frequency under massless scalar field perturbation is represented by a complex quantity which can be determined by the parameters of null circular orbits. The product of angular frequency ($\Omega_{c}$) with $l$  is defined as the real part whereas the product of Lyapunov exponent ($\lambda_{Null}$) with $(n+1/2)$  is called imaginary part of QNMs i.e.

\begin{equation}
\omega_{QNM}= l\Omega_{c}-i\left(n+\frac{1}{2}\right)\lambda_{Null},\label{QNMs}
\end{equation}
where  $n$ is the overtone number.

\noindent The above expression with Eqs.(\ref{eq47}) and (\ref{LYEXnull}) will yield the QNMs valid in eikonal regime of NCSBH  as follows 

	\begin{multline}
	\omega_{QNM} =l\sqrt{\frac{f(r_{c})}{r_{c}^{2}}}-i\left(n+\frac{1}{2}\right) \sqrt{\frac{f(r_{c})}{2  r_{c}^{2}}\left[\frac{24 M}{\sqrt{\pi}r_{c}}\left(\frac{\sqrt{\pi}}{2}-\Gamma \left(\frac{3}{2},\frac{r_{c}^{2}}{4\alpha}\right)\right)- \frac{e^{-\frac{r_{c}^{2}}{4\alpha}}M r_{c}^{3} \sqrt{\frac{r_{c}^{2}}{\alpha}}}{2\sqrt{\pi}\alpha^{2}}-4 \frac{e^{-\frac{r_{c}^{2}}{4\alpha}}M r_{c} \sqrt{\frac{r_{c}^{2}}{\alpha}}}{\sqrt{\pi}\alpha} -6 f(r_{c})\right]}.
	\end{multline}

\begin{figure}[H]
\centering
	\subfigure[]{\includegraphics[width=8cm,height=7.5cm]{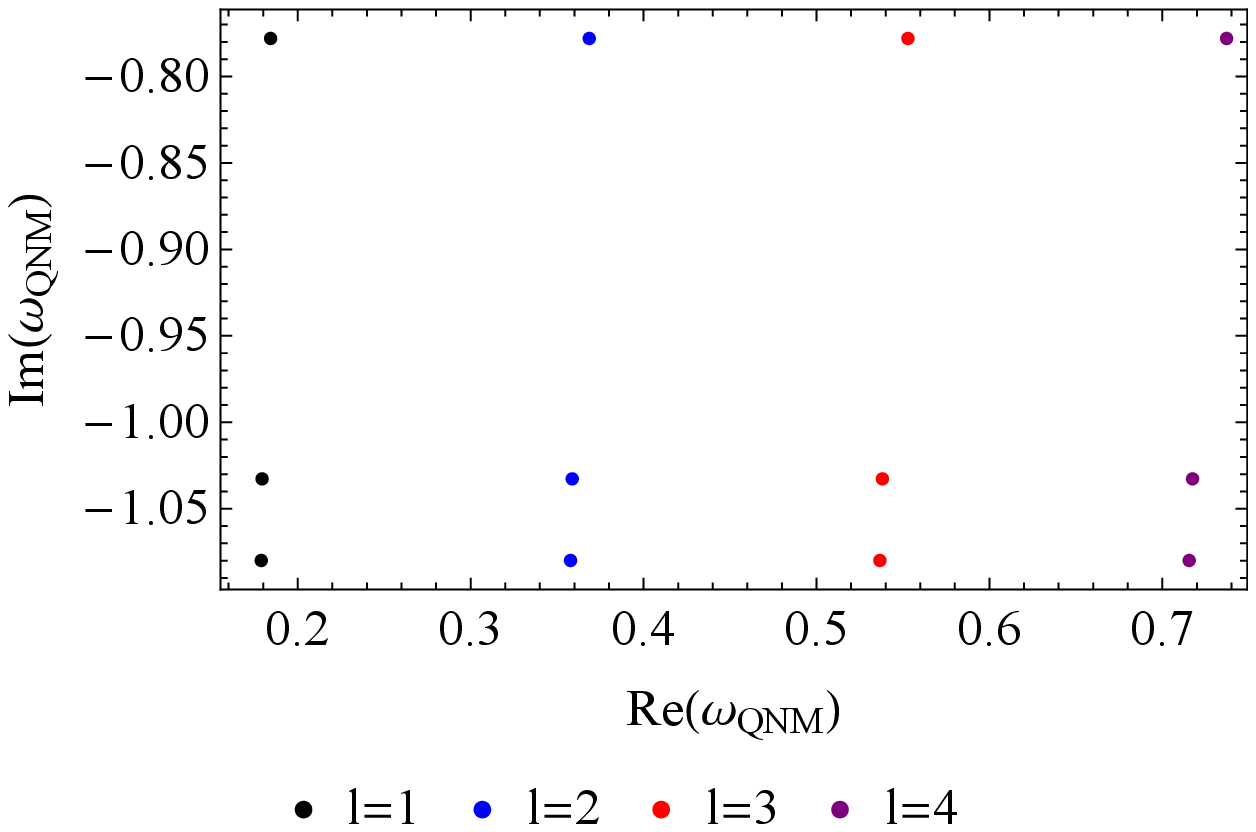}\label{QNM1}}
	\subfigure[]{\includegraphics[width=8cm,height=7.5cm]{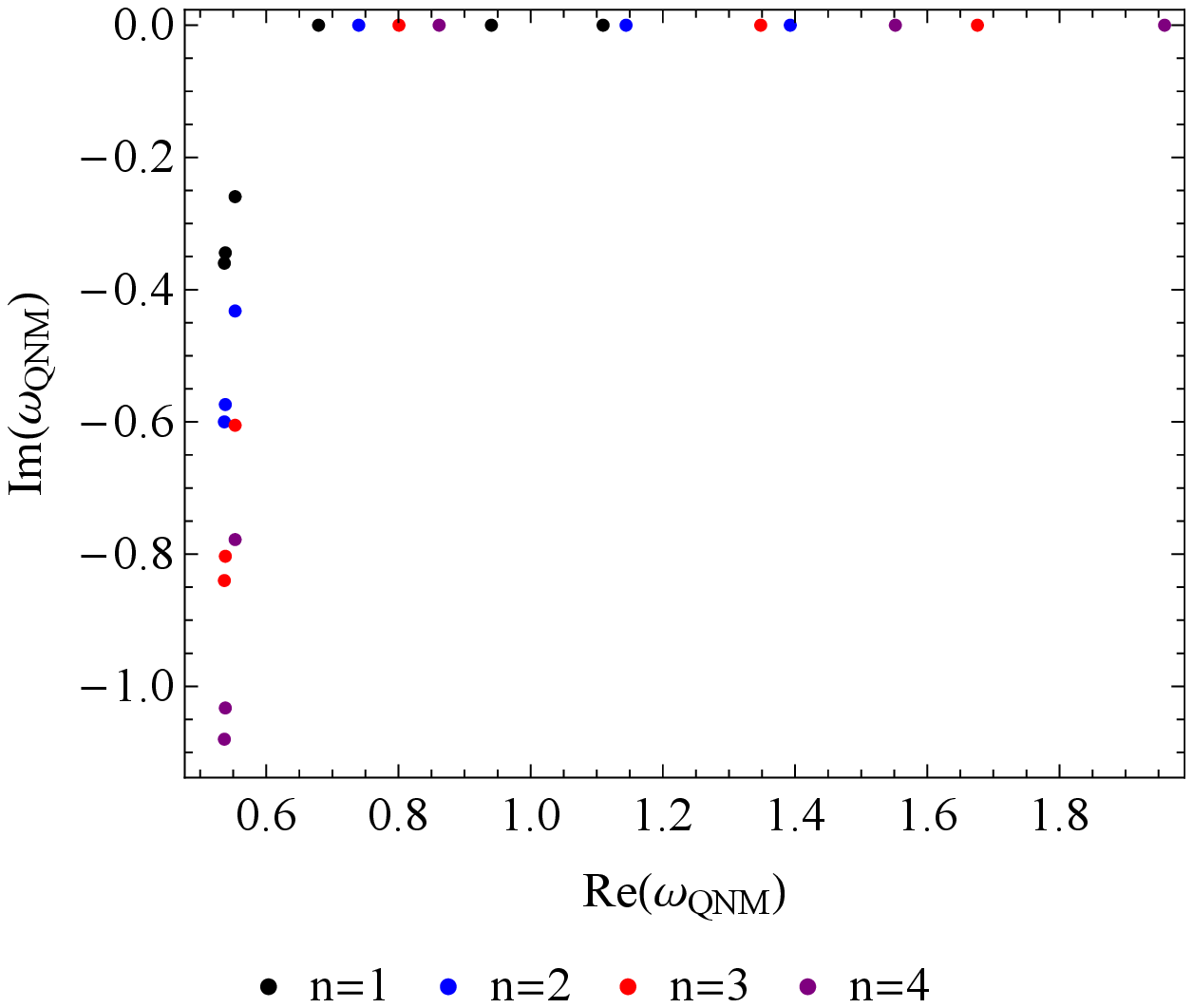}\label{QNM2}}
	\caption{The relation between real and imaginary parts of QNMs  $``\omega_{QNM}"$ by varying angular momentum of perturbation $l$ (see left panel) and by varying overtone number $n$ (see right panel) for fixed values of remaining parameters.}
\end{figure}
\noindent Significantly, one can interpret the Lyapunov exponent appearing in QNMs as the decay rate of unstable null circular orbits. It is obvious that the real and imaginary parts of QNMs of NCSBH depend on the non-commutative parameter ($\alpha$), therefore we intend to observe the effect of non-commutativity on QNM frequencies.\\
The complex list plot of QNMs i.e. the relation between the real and imaginary parts of the QNMs are visualised in \figurename{\ref{QNM1}} by varying angular momentum of perturbation $l$ by fixing remaining parameters. The same is presented in \figurename{\ref{QNM2}} by varying overtone number $n$ for fixed values of other parameters.\\

\noindent Further, we have evaluated the QNMs for the scalar field perturbation in the background of the NCSBH for different values of parameter $\alpha$  by varying perturbation parameter $l$ (listed in \tablename{~\ref{table1}}) and overtone number $n$ (listed in \tablename{~\ref{table2}}) for fixed values of other parameters.
\noindent Therefore, from \tablename{~\ref{table1}}, one can observe that  for fixed $l$, the real part of QNM increases and absolute value of the imaginary part of decreases as value of non-commutative parameter $\alpha$ increases.  Hence the oscillations damped more rapidly with increasing non-commutativity of spacetime. Similarly, from \tablename{~\ref{table2}}, it is observed that for fixed $n$, the real part of QNMs increases and absolute value of imaginary part decreases with increasing value of  $\alpha$ which represents rapid damping in oscillations.
\begin{table}[]
	\caption{QNMs of scalar field perturbation in the NCSBH spacetime by varying $l$ for fixed  $M = 1$,  $r_{c}= 2.5 $ and $ n = 4$.}
	\centering
	\begin{tabular}{|c|c|c|c|c|c|}
		\hline
		& 	$\omega_{QNM}(l=1)$ & 	$\omega_{QNM}(l=2)$ & 	$\omega_{QNM}(l=3)$ & 	$\omega_{QNM}(l=4)$ \\  [0.5ex]
		
		\hline
		$\alpha=0.1$ &	0.178886 - 1.0799 i & 0.357771 - 1.0799 i & 0.536657 - 1.0799  i & 0.715543 - 1.0799 i  \\
		\hline
		$\alpha=0.2$ &	0.179369 - 1.03271 i & 0.358738 - 1.03271   i & 0.538107 - 1.03271 i & 0.717476 - 1.03271  i \\
		\hline
		$\alpha=0.3$ &	0.184291 - 0.778012 i & 0.368582 - 0.778012  i & 0.552873 - 0.778012i & 0.737164 - 0.778012 i \\
		\hline
	\end{tabular} 
	\label{table1}
\end{table}		
\begin{table}[]
	\caption{QNMs of scalar field perturbation in the NCSBH spacetime by varying $n$ for fixed $M = 1$,  $r_{c}= 2.5 $ and $ l = 3$.}
	\centering
	\begin{tabular}{|c|c|c|c|c|c|}
		\hline
		& 	$\omega_{QNM}(n=1)$ & 	$\omega_{QNM}(n=2)$ & 	$\omega_{QNM}(n=3)$ & 	$\omega_{QNM}(n=4)$ \\  [0.5ex]
		\hline
		$\alpha=0.1$ &	0.536657\, - 0.359968 i & 0.536657\, - 0.599947 i & 0.536657\, - 0.839925 i & 0.536657\, - 1.0799 i \\
		\hline
		$\alpha=0.2$ &	0.538107\, - 0.344238 i & 0.538107\, - 0.57373 i & 0.538107\, - 0.803222 i & 0.538107\, - 1.03271 i  \\
		\hline
		$\alpha=0.3$ &	0.552873\, - 0.259337 i & 0.552873\, - 0.432229 i & 0.552873\, - 0.60512 i & 0.552873\, - 0.778012 i  \\
		
		\hline
		
	\end{tabular}  
	\label{table2}
\end{table}
\section{Summary and conclusions} 
The stability analysis of circular orbits of massive particles and photons around Schwarzschild BH with non-commutative geometry (NCSBH) has been investigated in detail. Also the QNMs of NCSBH has been derived and evaluated by employing massless scalar field perturbation in the eikonal approximation. Some of the interesting results drawn from our study are summarized as follows:
\begin{itemize}
	\item The standard SBH spacetime in the context of non-commutative inspired geometry turns out to have either two horizons (for $\alpha=0.1$), one horizon (for $\alpha=0.28$) and no horizon for higher value (for $\alpha=0.3$). Contrary to NCSBH spacetime geometry, there is only one horizon present in the case of SBH.  
	
	\item The respective plots of effective potential for NCSBH spacetime have concluded that the effective potential of such non-commutative geometry differs from the Schwarzschild case within the horizon radius ($r=2M$) of SBH as a consequence of the non-commutativity in spacetime. Beyond this limit of radius, all potentials coincide with each other conveying insignificant effect of non-commutativity.
	
	\item More specifically, we have derived the coordinate time Lyapunov exponent $(\lambda_{c})$ in order to study stability of timelike circular orbits on the equatorial plane of NCSBH spacetime. For different values of noncommutative parameter $\alpha$, the $\lambda_{c}$ decreases as radius of circular orbits increases and it shows the same behavior for all values of $\alpha$ beyond $r_{0}=3M$.
	
	\item In the case of null circular orbits, the radius of unstable circular orbits ($r_{c}$)  is found to be less than that the SBH case and depends upon the non-commutative parameter ($\alpha$).  So far, by deriving the instability exponent $\lambda_{Null}/\Omega_{c}$ for unstable circular orbits, it is observed that the instability increases to a highest value then decreases rapidly for circular orbits of higher radius. However, as the non-commutative parameter ($\alpha$) increases, the height of instability decreases accordingly and shows same behaviour for radius $r_{c}>3M$. The instability exponent versus $\alpha$ for circular orbit of $r_{c}=2.5$ shows that instability of null circular orbits decreases gradually with parameter $\alpha$.
	
	\item Further, we have employed the massless scalar field perturbation around NCSBH spacetime having unstable null circular orbits to compute the QNMs. In the eikonal limit i.e. for large $l$, the QNMs consist of a real part and a imaginary part depending upon orbital angular velocity $\Omega_{c}$ and the instability timescale of the orbit (i.e. Lyapunov exponent) for unstable null circular orbits respectively. In case of NCSBH, both parts of QNMs ($\omega_{QNM}$) are non-commutative parameter ($\alpha $) dependent. Furthermore, we have also calculated QNMs and discussed the relationship between real and imaginary parts by varying parameters $n$ and $l$ with increasing non-commutative parameter $\alpha$. It is concluded that the non-commutative parameter $\alpha$ has significant effect on damping in oscillations. The negative behavior of scalar field potential $V_{s}$ by varying parameters $\alpha$ and $l$ indicates instability of the null circular orbits of photons in NCSBH spacetime.
\end{itemize}

\section*{\normalsize Acknowledgments}
{\normalsize The authors would like to thank the anonymous referees
	for the constructive comments and suggestions which helped us to improve the presentation of this paper.  SG thankfully acknowledges the financial support provided by University Grants Commission (UGC), New Delhi, India as a Senior Research Fellow through UGC-Ref.No. {\bf 1479/CSIR-UGC NET-JUNE-2017}. HN would like to thank Science and Engineering Research Board (SERB), India for providing financial support through grant no. {\bf EMR/2017/000339}. The authors also acknowledge the facilities available at ICARD, Gurukula Kangri (Deemed to be University) Haridwar during period of this work.  }  

\bibliographystyle{unsrt} 
\bibliography{shobitrefLYEX} 
\end{document}